\documentclass[aps,pre,twocolumn,superscriptaddress,showpacs]{revtex4-2}

\usepackage{amsmath,amssymb,amsfonts}
\usepackage{natbib}
\usepackage{graphicx}
\usepackage{physics}
\usepackage{bm}
\usepackage{amsmath,amssymb,amsthm,amsfonts}
\usepackage{bbm}
\usepackage{hyperref}

\def\bibstyle#1{}
\def\bibdata#1{}
\makeatother

\begin{document}

\title{Geometric Entanglement Entropy on Projective Hilbert Space}

\author{Loris Di Cairano}
\affiliation{Department of Physics and Materials Science, University of Luxembourg, L-1511 Luxembourg City, Luxembourg}

\date{\today}

\begin{abstract}
Entanglement for pure bipartite states is most commonly quantified in a state-by-state manner to each pure state of a bipartite system a scalar quantity, such as the von Neumann entropy of a reduced density matrix. This provides a precise local characterization of how entangled a given state is. At the same time, this local description naturally invites a set of complementary, more global questions about the structure of the space of pure states: How abundant are the states with a given amount of entanglement within the full state space? Do the manifolds of constant entanglement exhibit distinct geometric regimes? These questions shift the focus from assigning an entanglement value to a single state to understanding the global
organization and geometry of entanglement across the entire manifold of pure states.

In this work, we develop a geometric framework in which these questions become natural. We regard the projective Hilbert space of pure states, endowed with the Fubini–Study metric, as a Riemannian manifold and promote bipartite entanglement to a macroscopic functional on this manifold. Its level sets stratify the space of pure states into hypersurfaces of constant entanglement, and we define a geometric entanglement entropy as the log-volume of these hypersurfaces, weighted by the Fubini–Study gradient of entanglement. This quantity plays the role of a microcanonical entropy in ``entanglement space'': it measures the degeneracy of a given entanglement value in the natural quantum geometry.

The framework is illustrated first in the simplest case of a single spin-$1/2$ (where the Fubini–Study metric reduces to the Bloch sphere
metric) and then for bipartite entanglement of spin systems, including a two-qubit example where explicit calculations can be carried out, along with a sketch of the extension to spin chains.
\end{abstract}

\maketitle

\section{Introduction}

Quantum information theory has developed a rich set of tools to quantify and manipulate entanglement~\cite{nielsen_chuang,horodecki_review}. For pure bipartite states, the standard measure is the von Neumann entropy of the reduced density matrix, which assigns to each state a single real number between zero and its maximum value. This approach is intrinsically local in the space of states: it asks, for a given wavefunction, ``how entangled is this state?''

This local viewpoint thus stimulates several questions about global properties: \emph{How abundant or rare are states with a given entanglement within the set of all pure states?} \emph{Does the space of states admit distinct geometric regimes when organized according to entanglement, suggesting entanglement ``phases''?} \emph{How rigid is entanglement under infinitesimal deformations of the wavefunction?} These are questions about the geometry and measure-theoretic structure of the space of pure states, which play an important role in our understanding of typical entanglement in large systems~\cite{page1993typical,foong_kanno,typical_entanglement_review,popescu2006entanglement,hayden2006aspects}, entanglement in many-body systems and quantum field theory~\cite{Eiser2008entanglement,eisert2010colloquium,calabrese2009entanglement}, and the geometry of quantum state space~\cite{bengtsson_zyczkowski,provost1980riemannian,brody2001geometric}. They are also central in current studies of entanglement growth and entanglement phase transitions in random quantum circuits and monitored dynamics~\cite{nahum2017quantum,li2018quantum,skinner2019measurement}, as well as in the design of highly entangled resource states for quantum technologies.

In this work, we propose a geometric framework that addresses such questions at the level of kinematics, without assuming any specific Hamiltonian or dynamical model. We regard the projective Hilbert space of pure states as a Riemannian manifold, endowed with the Fubini–Study metric. Bipartite entanglement is promoted to a macroscopic functional on this manifold; its level sets stratify the state space into hypersurfaces of constant entanglement. This allows us to define a \emph{geometric entanglement entropy}, which measures, in the Fubini–Study geometry, how large the set of pure states is that realizes a given entanglement value.

The central idea is analogous to microcanonical constructions in classical statistical mechanics \cite{di2022geometrictheory,di2021topology,franzosi2018microcanonical,franzosi2019microcanonical,bel2020geometrical,gori_configurational,gori2022topological}: there, energy is treated as a macroscopic variable, and entropy is the logarithm of the volume of the corresponding energy shell in phase space. Here, entanglement plays the role of the macroscopic variable, the projective Hilbert space with the Fubini–Study metric plays the role of phase space, and the constant-entanglement hypersurfaces take the role of energy shells. The resulting entropy function encodes the abundance of different entanglement values and relates its derivatives to the curvature invariants of the hypersurfaces that carry the entanglement structure.

Our work can be regarded as a quantum-information analog of this program: we replace the Hamiltonian energy with a bipartite entanglement functional on projective Hilbert space and construct a microcanonical-like entropy that counts the Fubini–Study volume of constant-entanglement manifolds.

Geometric approaches to entanglement are not rare but are already present in the literature, 
including geometric entanglement measures based on distances to separable states and approaches 
that exploit the geometry of state space and reduced density matrices ~\cite{wei_geometric_entanglement,orus_geometric_entanglement,leinaas_geometrical_aspects,frydryszak_geometric_measure,brody2001geometric}. More recently, the Fubini–Study metric and related Riemannian structures on projective Hilbert space have been 
used to define entanglement distances and geometric quantum entropies ~\cite{vesperini_geometric_entanglement,anza_maximum_geometric_entropy,qid_geometric_entropy}. 

In particular, a series of works by Franzosi and collaborators has exploited the Fubini–Study 
geometry to construct entanglement measures that can be applied to a broad class of systems. 
An entanglement distance derived from the Fubini–Study metric was introduced for arbitrary 
$M$-qudit hybrid systems \cite{cocchiarella2020entanglement}, and then 
extended to multipartite mixed states as a unified framework for entanglement and more general 
quantum correlations \cite{vesperini2023entanglement}. 
The same geometric distance has been applied to graph-based architectures, both in undirected 
graph states \cite{vesperini2024entanglement} and in quantum 
systems defined on directed graphs \cite{de2025entanglement,de2025entanglement}, where it quantitatively relates multipartite entanglement to 
network connectivity features.

All these contributions make systematic use of the Fubini–Study metric to assign entanglement 
and correlation measures to individual quantum states, and show that this geometry provides a 
natural language to capture nonclassical correlations in a variety of architectures. The present work is complementary in spirit: instead of defining a sensible state-by-state entanglement 
measure, we treat entanglement as a macroscopic scalar functional on projective Hilbert space and construct a microcanonical-like entropy that counts the Fubini–Study volume of the constant-entanglement hypersurfaces.

The goals of the paper are threefold:
\begin{itemize}
  \item to formulate a clean, model-independent geometric framework for entanglement on projective Hilbert space;
  \item to define a geometric entanglement entropy that captures the degeneracy of entanglement values in the natural quantum geometry;
  \item to outline how curvature and Fubini-Study quantities encode the rigidity and possible ``phases'' of entanglement, and to illustrate the formalism in simple spin systems.
\end{itemize}

\section{Projective Hilbert space and Fubini-Study geometry}
\label{sec:geometry}

Let $\mathcal H$ be a finite-dimensional complex Hilbert space of dimension $d$, for
instance a spin system $(\mathbb C^2)^{\otimes N}$ of $N$ qubits. Pure states are rays
in $\mathcal H$: equivalence classes $[\psi]$ of non-zero vectors $|\psi\rangle$ under
nonzero complex rescalings,
\begin{equation}
  [\psi] := \{\lambda|\psi\rangle : \lambda\in\mathbb C\setminus\{0\}\}.
\end{equation}
The set of all rays forms the projective Hilbert space
$\mathcal P(\mathcal H) \simeq \mathbb{CP}^{d-1}$, which is a compact complex manifold.

Given a normalized representative $|\psi\rangle$ with $\langle\psi|\psi\rangle=1$, a
tangent vector at $[\psi]$ can be represented by a variation $|\delta\psi\rangle$
in $\mathcal H$ modulo changes that amount only to a change of global phase.
A convenient choice is to impose the orthogonality condition
\begin{equation}
  \langle\psi|\delta\psi\rangle = 0,
  \label{eq:tangent_cond}
\end{equation}
which selects a unique horizontal representative of the tangent vector.

The Fubini-Study metric $g_{\mathrm{FS}}$ is the unique (up to scale) unitarily
invariant K\"ahler metric on $\mathcal P(\mathcal H)$~\cite{bengtsson_zyczkowski}. In terms
of a normalized representative and an arbitrary variation $|\delta\psi\rangle$, the
corresponding squared line element is
\begin{equation}
  ds^2_{\mathrm{FS}} = 4\left( \langle\delta\psi|\delta\psi\rangle
  - |\langle\psi|\delta\psi\rangle|^2 \right).
  \label{eq:fs_general}
\end{equation}
Imposing the tangency condition~\eqref{eq:tangent_cond} simplifies this to
\begin{equation}
  ds^2_{\mathrm{FS}} = 4\,\langle\delta\psi|\delta\psi\rangle.
  \label{eq:fs_tangent}
\end{equation}
Thus the norm of a tangent vector $|\delta\psi\rangle$ in the Fubini-Study metric is
proportional to the standard Hilbert space norm of the orthogonal variation.

The metric $g_{\mathrm{FS}}$ induces a unitarily invariant volume form $d\mu_{\mathrm{FS}}$
on $\mathcal P(\mathcal H)$, which coincides with the Haar measure on pure states
used in studies of typical entanglement~\cite{page1993typical,bengtsson_zyczkowski}.
We shall view $(\mathcal P(\mathcal H),g_{\mathrm{FS}})$ as the quantum analogue of
a classical phase space endowed with a Riemannian metric.

\subsection{Entanglement as a scalar functional}

We now fix a bipartition of the Hilbert space,
\begin{equation}
  \mathcal H = \mathcal H_A \otimes \mathcal H_B,
\end{equation}
with dimensions $d_A = \dim\mathcal H_A$ and $d_B = \dim\mathcal H_B$, and consider
pure states $|\psi\rangle_{AB} \in \mathcal H$. For each such state, the reduced density
matrix of subsystem $A$ is
\begin{equation}
  \rho_A(\psi) = \mathrm{Tr}_B\,|\psi\rangle\langle\psi|.
\end{equation}
The bipartite entanglement entropy of $|\psi\rangle_{AB}$ is defined as the von Neumann
entropy of $\rho_A$~\cite{nielsen_chuang,horodecki_review},
\begin{equation}
  E([\psi]) = S(\rho_A(\psi))
  = -\mathrm{Tr}\big(\rho_A(\psi)\log\rho_A(\psi)\big),
\end{equation}
where we regard $E$ as a function of the ray $[\psi] \in \mathcal P(\mathcal H)$.
Thus we have a smooth scalar functional
\begin{equation}
  E: \mathcal P(\mathcal H)\to [0,\log d_A],
\end{equation}
which assigns to each pure state its bipartite entanglement between $A$ and $B$.

The level sets of $E$,
\begin{equation}
  \Sigma_e := \{[\psi]\in\mathcal P(\mathcal H): E([\psi])=e\},
\end{equation}
define a stratification of the projective Hilbert space into manifolds of constant
entanglement. For regular values of $e$, each $\Sigma_e$ is a smooth hypersurface
of codimension one in $\mathcal P(\mathcal H)$.

The gradient of $E$ with respect to the Fubini-Study metric is the vector field
$\nabla^{\mathrm{FS}} E$ defined implicitly by
\begin{equation}
  g_{\mathrm{FS}}(\nabla^{\mathrm{FS}}E, X) = dE(X)
  \qquad \text{for all } X \in T\mathcal P(\mathcal H).
\end{equation}
Geometrically, $\nabla^{\mathrm{FS}}E$ is orthogonal to the level sets $\Sigma_e$ and
points in the direction of maximal increase of $E$ per unit Fubini-Study distance.
Its norm $\|\nabla^{\mathrm{FS}}E\|_{g_{\mathrm{FS}}}$ quantifies the local susceptibility
of entanglement to infinitesimal deformations of the wavefunction.

We shall denote by
\begin{equation}
  \bm \xi = \frac{\nabla^{\mathrm{FS}}E}{\|\nabla^{\mathrm{FS}}E\|^2}
\end{equation}
the vector field orthogonal to the constant-entanglement hypersurfaces $\Sigma_e$ generates the flow along increasing values of $e$, i.e., $dE(\bm\xi)=1$ \cite{hirsch2012differential}.\\

We now make explicit the coordinate form of the Fubini-Study metric and its
decomposition along and orthogonal to a given scalar functional $E$ (for instance,
the entanglement entropy).

Let $[\psi]\in\mathcal P(\mathcal H)$ and choose local coordinates
$x^\mu$ on a patch of projective Hilbert space, so that normalized states
are written as
\begin{equation}
    |\psi(x)\rangle, \qquad x = (x^1,\dots,x^n).
\end{equation}
A tangent vector is represented by a variation
\begin{equation}
    |\delta\psi\rangle
    = \partial_\mu|\psi(x)\rangle\,\delta x^\mu,
\end{equation}
where $\partial_\mu:=\partial/\partial x^\mu$ and $\delta x^\mu$ are the
components of the tangent vector in the coordinate basis.

The Fubini-Study line element can be written in the usual Riemannian form
\begin{equation}
    ds^2_{\mathrm{FS}} = g_{\mu\nu}(x)\,dx^\mu dx^\nu,
\end{equation}
with components
\begin{equation}
    g_{\mu\nu}(x)
    = 4\,\mathrm{Re}\!\left[
        \langle\partial_\mu\psi|\partial_\nu\psi\rangle
        - \langle\partial_\mu\psi|\psi\rangle\,
          \langle\psi|\partial_\nu\psi\rangle
      \right].
    \label{eq:FS_metric_components}
\end{equation}
If we choose the ``horizontal'' gauge $\langle\psi|\partial_\mu\psi\rangle=0$
for all $\mu$, this simplifies to
\begin{equation}
    g_{\mu\nu}(x)
    = 4\,\mathrm{Re}\langle\partial_\mu\psi|\partial_\nu\psi\rangle,
\end{equation}
and the squared norm of a tangent vector is
\begin{equation}
    ds^2_{\mathrm{FS}}
    = g_{\mu\nu}\,\delta x^\mu\delta x^\nu
    = 4\,\langle\delta\psi|\delta\psi\rangle,
\end{equation}
which is precisely Eq.~\eqref{eq:fs_tangent}. In other words,
Eq.~\eqref{eq:fs_tangent} is the same Fubini-Study metric written in
components in a coordinate system $\xi^\mu$ and in the gauge
$\langle\psi|\partial_\mu\psi\rangle=0$.

As a concrete example, consider a single qubit, with
\begin{equation}
    |\psi(\theta,\phi)\rangle
    = \cos\frac{\theta}{2}\,|\!\uparrow\rangle
      + e^{i\phi}\sin\frac{\theta}{2}\,|\!\downarrow\rangle,
\end{equation}
so that $\xi^1=\theta$, $\xi^2=\phi$. A straightforward computation of
\eqref{eq:FS_metric_components} gives
\begin{equation}
    ds^2_{\mathrm{FS}}
    = d\theta^2 + \sin^2\theta\,d\phi^2,
\end{equation}
i.e.\ $g_{\theta\theta}=1$, $g_{\phi\phi}=\sin^2\theta$, $g_{\theta\phi}=0$,
the standard round metric on the Bloch sphere.

\medskip

Let now $E:\mathcal P(\mathcal H)\to\mathbb R$ be a smooth scalar functional,
and denote by $\partial_\mu E$ its coordinate derivatives and by
$g^{\mu\nu}$ the inverse of $g_{\mu\nu}$. The gradient of $E$ with respect
to the Fubini-Study metric has components
\begin{equation}
    (\nabla E)^\mu = g^{\mu\nu}\,\partial_\nu E,
\end{equation}
and squared norm
\begin{equation}
    \|\nabla E\|^2
    = g_{\mu\nu}(\nabla E)^\mu(\nabla E)^\nu.
\end{equation}
We define the covariant components of the unit normal to the level sets
$\Sigma_e=\{x:E(x)=e\}$ by
\begin{equation}
    n_\mu
    := \frac{\partial_\mu E}{\|\nabla E\|},
    \qquad
    n_\mu n_\nu g^{\mu\nu} = 1.
\end{equation}
In terms of $n_\mu$ the Fubini-Study metric can be decomposed into a
normal and a tangential part:
\begin{equation}
    g_{\mu\nu}(x)
    = n_\mu n_\nu + \sigma_{\mu\nu}(x),
    \label{eq:g_decomposition}
\end{equation}
so that the line element reads
\begin{equation}
    ds^2_{\mathrm{FS}}
    = \frac{\partial_\mu E\,\partial_\nu E}{\|\nabla E\|^2}\,
      dx^\mu dx^\nu
      + \sigma_{\mu\nu}(x)\,dx^\mu dx^\nu.
    \label{eq:FS_split}
\end{equation}
The first term in~\eqref{eq:FS_split} measures displacements in the
direction orthogonal to the constant--$E$ hypersurfaces, while the second term is the induced metric on the level sets $\Sigma_e$. By construction,
$\sigma_{\mu\nu}$ is purely tangential,
\begin{equation}
    \sigma_{\mu\nu}(x)\,g^{\nu\rho}(x)\,\partial_\rho E(x) = 0,
\end{equation}
so that $\sigma_{\mu\nu}$ encodes the intrinsic geometry of the
constant--entanglement manifolds, whereas the normal piece
$n_\mu n_\nu$ singles out the direction along which $E$ changes.

Finally, noting that $dE=\partial_\mu E\,dx^\mu$, the metric takes the form~\cite{zhou2013simple,di2022geometrictheory}
\begin{equation}
    g_{\mathrm{FS}}
    = \frac{dE}{\|\nabla E\|}\,
      \otimes \frac{dE}{\|\nabla E\|}
      + \sigma_{\mu\nu}(x)\,dx^\mu \otimes dx^\nu.
    \label{eq:FS_standard}
\end{equation}

Equations~\eqref{eq:FS_metric_components} and~\eqref{eq:g_decomposition}
describe the same Fubini-Study metric: the former gives its components
in a coordinate basis $\{\xi^\mu\}$, the latter rearranges those components
into a contribution along $\nabla E$ plus a contribution tangent to the
hypersurfaces $E=\text{const}$.

\section{Geometric entanglement entropy}
\label{sec:geo_entropy}

The functional $E([\psi])$ assigns a macroscopic quantity (entanglement) to each
point of the state manifold. We now ask: \emph{how large is, in the Fubini-Study
geometry, the set of states with a given entanglement value $e$?} This leads to
the following microcanonical-type construction.

\subsection{Density of states at fixed entanglement}

We define the \emph{density of states at fixed entanglement} as
\begin{equation}
    \omega(e)=\int \delta(E(|\psi|-e))\,d\mu_{\rm{FS}}(|\psi|)\,.
\end{equation}
Through the coarea formula~\cite{federer_gmt}, it can be written as
\begin{equation}
  \omega(e) := \int_{\Sigma_e}
  \frac{d\sigma_{\mathrm{FS}}}{\|\nabla^{\mathrm{FS}}E\|_{g_{\mathrm{FS}}}},
  \label{eq:omega_def}
\end{equation}
where $d\sigma_{\mathrm{FS}}$ is the volume element induced on $\Sigma_e$ by the
Fubini-Study metric and the denominator takes into account the Jacobian relating the
Fubini-Study volume of a thin layer around $\Sigma_e$ to the volume of the
hypersurface itself. Equation~\eqref{eq:omega_def} is the direct analog of the
classical microcanonical density of states on energy shells, with entanglement
playing the role of energy and projective Hilbert space replacing phase space.

Operationally, the density $\omega(e)$ can be interpreted as follows. Sampling pure
states with the unitarily invariant measure $d\mu_{\mathrm{FS}}$, the probability
density $P(e)$ to observe a value of entanglement between $e$ and $e+de$ satisfies
\begin{equation}
  P(e)\,de \propto \omega(e)\,de.
\end{equation}
Thus $\omega(e)$ quantifies the \emph{degeneracy} of the entanglement value $e$:
the effective number of distinct wavefunctions that realize that amount of entanglement in the natural quantum geometry.

\subsection{Definition and interpretation of $S_{\mathrm{geo}}(e)$}

We now define the \emph{geometric entanglement entropy} as
\begin{equation}
  S_{\mathrm{geo}}(e) := \log\omega(e),
\end{equation}
up to an arbitrary additive constant. This quantity is conceptually distinct from
the von Neumann entanglement entropy $E([\psi])$ of a single state. While $E([\psi])$
measures ``how entangled'' a given wavefunction is, the geometric entropy
$S_{\mathrm{geo}}(e)$ measures ``how many'' pure states are consistent with the
same entanglement value $e$.

In other words, $S_{\mathrm{geo}}(e)$ plays the role of microcanonical entropy
in entanglement space: it is a function of a macroscopic variable $e$ that counts the Fubini–Study volume of the manifold of microstates (pure wavefunctions)
compatible with that value of entanglement. Large values of $S_{\mathrm{geo}}(e)$
indicate entanglement regimes that are geometrically abundant; small values correspond to entanglement regimes that are intrinsically rare in the space of pure states.

This perspective complements the standard state-by-state quantification of
entanglement~\cite{vedralReview,horodecki_review}. It provides a geometric notion of the \emph{typicality} of different entanglement values, formulated directly in terms of the structure of projective Hilbert space, and it does so in a way that is
independent of any particular Hamiltonian or physical model.

\section{Geometry of entanglement level sets}
\label{sec:flow}

The density of states $\omega(e)$ and the corresponding geometric entropy
$S_{\mathrm{geo}}(e)$ capture how the Fubini-Study volume of the constant-entanglement
hypersurfaces $\Sigma_e$ depends on $e$. It is natural to ask how the derivatives
$S_{\mathrm{geo}}'(e)$ and $S_{\mathrm{geo}}''(e)$ are related to the geometric properties of the hypersurfaces themselves, such as mean and scalar curvature.

The family $\{\Sigma_e\}$ forms a smooth foliation of $\mathcal P(\mathcal H)$ (except possibly at singular values of $E$). The vector field $\bm \xi$ defines the second fundamental form and the Weingarten operator $W_e : T\Sigma_e \to T\Sigma_e$.
Their traces and powers,
\begin{equation}
  H_{\bm\xi} = \mathrm{Tr}\,W_{\bm\xi}=\text{div}^{\text{FS}}\,\bm\xi, 
\end{equation}
encode the mean curvature. 

By following the same approach as in Ref.~\cite{gori_configurational,di2022geometrictheory,di2021topology}, we can write
\begin{equation}\label{def:der-S}
\begin{split}
        \partial_e S_{\text{geo}}(e)&=\frac{1}{\omega(e)}\int_{\Sigma_e}\text{Tr}[W_{\bm\xi}]\frac{d\sigma_{\mathrm{FS}}}{\|\nabla^{\mathrm{FS}}E\|_{g_{\mathrm{FS}}}}\\
        &=\int_{\Sigma_e}\text{Tr}[W_{\bm\xi}]~d\rho_{\mathrm{FS}}
\end{split}
\end{equation}
where $d\rho_{\text{FS}}:=\omega(e)^{-1}d\sigma_{\mathrm{FS}}/\|\nabla^{\mathrm{FS}}E\|_{g_{\mathrm{FS}}}$, which plays the role of a statistical measure, indeed 
\[
    \int_{\Sigma_e}d\rho_{\text{FS}}=\omega(e)^{-1}\int_{\Sigma_e}\frac{d\sigma_{\mathrm{FS}}}{\|\nabla^{\mathrm{FS}}E\|_{g_{\mathrm{FS}}}}=1
\]


This construction admits a clear geometric and physical interpretation in the present context. The quantity $\omega(e)$ measures the Fubini-Study volume of the constant-entanglement manifold $\Sigma_e$, so that $S_{\mathrm{geo}}(e)=\log\omega(e)$ plays the role of a microcanonical entropy in entanglement space. The normalized measure $d\rho_{\mathrm{FS}}$ is then the natural microcanonical distribution over pure states with fixed entanglement $e$. Eq.~\eqref{def:der-S} shows that the \emph{slope} of the geometric entanglement entropy is the
microcanonical average, over all states with entanglement $e$, of the trace of the Weingarten map $W_{\bm\xi}$---i.e., of the mean extrinsic curvature of $\Sigma_e$
along the normal direction $\bm\xi \propto \nabla^{\mathrm{FS}}E$. Physically, this means that $\partial_e S_{\mathrm{geo}}(e)$ quantifies how rapidly the Fubini-Study volume of the constant-entanglement shell changes when one moves to a neighboring shell at slightly different entanglement: large positive values correspond to
entanglement regimes where an infinitesimal increase of $e$ opens up a rapidly growing family of states, whereas small values indicate a geometrically ``rigid''
entanglement sector, in which the available Fubini-Study volume grows only slowly with $e$. \\

We now illustrate these constructions in simple spin systems.

\section{Examples}
\label{sec:examples}

\subsection{Single spin-$1/2$ and the Bloch sphere}

As a warm-up, consider a single spin-$1/2$. In this case the Hilbert space is
$\mathcal H = \mathbb C^2$, and the projective Hilbert space
$\mathcal P(\mathcal H)\simeq\mathbb{CP}^{1}$ is isomorphic to the Bloch sphere
$S^2$~\cite{bengtsson_zyczkowski}. A normalized spinor can be parametrized as
\begin{equation}
  |\psi(\theta,\phi)\rangle
  = \cos\frac\theta2\,|\uparrow\rangle
  + e^{i\phi}\sin\frac\theta2\,|\downarrow\rangle,
\end{equation}
with $(\theta,\phi)\in[0,\pi]\times[0,2\pi)$.

The Fubini-Study metric becomes the standard round metric on the sphere,
\begin{equation}
  ds^2_{\mathrm{FS}} = d\theta^2 + \sin^2\theta\,d\phi^2.
\end{equation}
In this single-spin example there is no nontrivial bipartite entanglement:
any bipartition would isolate a pure state on one side and a trivial system
on the other, and the entanglement entropy vanishes identically. Nonetheless,
this example shows how the Fubini-Study geometry reduces to the familiar
Bloch sphere structure for a single qubit, and it provides a simple arena for
visualizing geodesics, distances and volumes in state space.

\subsection{Two-qubit entanglement}

We now consider the simplest nontrivial bipartite system: two spin-$1/2$
subsystems, with $\mathcal H_A\simeq\mathbb C^2$, $\mathcal H_B\simeq\mathbb C^2$,
and $\mathcal H = \mathcal H_A\otimes\mathcal H_B$. Any pure state
$|\psi\rangle\in\mathcal H$ can be written in Schmidt form as
\begin{equation}
  |\psi(\theta)\rangle
  = \cos\theta\,|00\rangle + \sin\theta\,|11\rangle,
  \qquad 0\le \theta\le\frac\pi4,
  \label{eq:two_qubit_schmidt}
\end{equation}
up to local unitaries. The reduced density matrix of subsystem $A$ is
\begin{equation}
  \rho_A(\theta) = 
  \begin{pmatrix}
    \cos^2\theta & 0 \\
    0 & \sin^2\theta
  \end{pmatrix},
\end{equation}
and the entanglement entropy is
\begin{equation}
  E(\theta) = -\cos^2\theta \log\cos^2\theta
              -\sin^2\theta\log\sin^2\theta,
\end{equation}
which is simply the binary entropy of $\cos^2\theta$. It varies monotonically
from $E(0)=0$ (product state) to $E(\pi/4)=\log 2$ (maximally entangled Bell state).

In this reduced description, parametrized by the single angle $\theta$,
the relevant portion of projective Hilbert space is effectively one-dimensional:
we are following a curve of states modulo local unitaries, which form equivalence
classes under which entanglement is invariant~\cite{vedralReview}. Along this curve,
the Fubini-Study line element can be written as
\begin{equation}
  ds^2_{\mathrm{FS}} = 4\,\langle\partial_\theta\psi|
  \partial_\theta\psi\rangle\,d\theta^2
  = 4\left( \cos^2\theta + \sin^2\theta \right)d\theta^2
  = 4\,d\theta^2,
\end{equation}
so that the Fubini-Study distance along the curve is proportional to $|\Delta\theta|$.

The level sets of $E$ in the full projective space $\mathcal P(\mathcal H)$ are
three-dimensional hypersurfaces (since $\mathcal P(\mathcal H)\simeq\mathbb{CP}^3$
has real dimension six). For each fixed $\theta$ they consist of all pure
two-qubit states that share the same Schmidt coefficients $(\cos\theta,\sin\theta)$.
Their geometry is nontrivial, but many global properties can already be seen
in the Schmidt-reduced picture: the entanglement is maximal at $\theta=\pi/4$, where
a small variation of $\theta$ produces only a small change in $E(\theta)$,
and vanishes at the endpoints, where the gradient of $E$ with respect to $\theta$
also vanishes. Between these extremes, $E(\theta)$ has a single inflection point,
which reflects a change in the local curvature of the constant-entanglement
hypersurfaces in the full state space.

In principle, one could compute the density of states $\omega(e)$ by sampling
pure two-qubit states with the Fubini-Study measure, computing their
entanglement entropy, and constructing an empirical distribution. This would
yield the geometric entanglement entropy $S_{\mathrm{geo}}(e)$ in this system,
which in turn encodes how the Fubini-Study volume of the constant-entanglement
hypersurfaces varies between product states and maximally entangled states.
In this simple example $S_{\mathrm{geo}}(e)$ can be obtained analytically or
semi-analytically by exploiting the known distribution of Schmidt coefficients
for random two-qubit states~\cite{bengtsson_zyczkowski}.

\subsection{Spin chains and bipartite entanglement}

We now sketch the extension of the framework to many-body spin systems,
such as chains of $N$ spin-$1/2$ sites. The Hilbert space is
$\mathcal H = (\mathbb C^2)^{\otimes N}$, and we consider a bipartition into a
block $A$ of $\ell$ spins and its complement $B$ of $N-\ell$ spins,
\begin{equation}
  \mathcal H = \mathcal H_A\otimes\mathcal H_B,
\end{equation}
with dimensions $d_A=2^\ell$, $d_B=2^{N-\ell}$. For each pure many-body state
$|\psi\rangle$ we define the entanglement entropy $E([\psi]) = S(\rho_A(\psi))$
as before, and interpret it as a scalar functional on the projective space
$\mathcal P(\mathcal H)$.

The constant-entanglement hypersurfaces $\Sigma_e$ now live in a manifold of
dimension $2^{N+1}-2$ (real), and their geometry is extremely rich. Nevertheless,
the construction of the geometric entanglement entropy proceeds exactly as in
lower-dimensional examples:
\begin{itemize}
  \item The Fubini-Study metric $g_{\mathrm{FS}}$ is defined by
  Eq.~\eqref{eq:fs_general}, and induces a volume form $d\mu_{\mathrm{FS}}$.
  \item The entanglement entropy $E([\psi])$ defines level sets
  $\Sigma_e = \{[\psi]:E([\psi])=e\}$.
  \item The gradient $\nabla^{\mathrm{FS}}E$ and its norm quantify the local
  sensitivity of entanglement to changes in the wavefunction.
  \item The density of states at fixed entanglement $\omega(e)$ is given by
  Eq.~\eqref{eq:omega_def}, and the geometric entropy
  $S_{\mathrm{geo}}(e) = \log\omega(e)$ measures the Fubini-Study volume
  of the manifold of states with entanglement $e$.
\end{itemize}

In this many-body context, $S_{\mathrm{geo}}(e)$ encodes how different regimes
of bipartite entanglement (area-law, logarithmic scaling, volume-law) are
represented in the full space of many-body wavefunctions. For instance, one
expects that states with volume-law entanglement occupy an overwhelming portion
of the Fubini-Study volume for large $N$~\cite{page1993typical,bengtsson_zyczkowski},
whereas states with strict area-law entanglement form a manifold of much smaller
volume, associated with ground states of local Hamiltonians and matrix product
states~\cite{EisertReview}. The geometric entanglement entropy $S_{\mathrm{geo}}(e)$
provides a quantitative way to express this intuition: it is the function that
measures the logarithmic volume of the constant-entanglement hypersurfaces.

This opens the possibility of defining entanglement ``phases'' directly in the space of many-body wavefunctions, without reference to a specific Hamiltonian: different curvature regimes of $\Sigma_e$ (e.g., regions where certain averaged curvatures are positive, negative, or vanish) correspond to qualitatively distinct
structures of the set of states with fixed entanglement. Transitions between these regimes, signaled by changes in the behavior of $S_{\mathrm{geo}}(e)$, could be interpreted as geometric entanglement transitions in state space.

\section{Illustrative computation of the Weingarten trace}
\label{sec:weingarten_examples}

In the previous section, we defined the constant--entanglement hypersurfaces \(\Sigma_e = \{[\psi]:E([\psi])=e\}\) in projective Hilbert space and introduced the vector field
\begin{equation}
  \bm \xi := \frac{\nabla^{\mathrm{FS}} E}{\|\nabla^{\mathrm{FS}}E\|_{g_{\mathrm{FS}}}^2}.
\end{equation}
In the geometric microcanonical construction, the trace of the Weingarten operator (the mean curvature in the chosen convention) naturally appears through the divergence of this field. More precisely, in local coordinates
on \(\mathcal P(\mathcal H)\) one considers
\begin{equation}
  \mathrm{Tr}\,W_{\bm\xi}= \mathrm{div}^{\text{FS}}\,\bm \xi
  = \frac{1}{\sqrt{\det g}}\,
    \partial_i\big(\sqrt{\det g}\,\xi^i\big),
  \label{eq:div_general}
\end{equation}
where \(g\) is the Fubini-Study metric and \(\xi^i\) are the contravariant components of \(\bm \xi\) in those coordinates. In this section, we illustrate how this works in simple cases where explicit calculations can be carried out.

\subsection{Warm-up: a single qubit on the Bloch sphere}

Although a single qubit does not support nontrivial bipartite entanglement,
it provides the simplest arena to observe the geometry at work. The Hilbert space is \(\mathcal H=\mathbb C^2\) and the projective Hilbert space \(\mathcal P(\mathcal H)\simeq\mathbb{CP}^1\) is the Bloch sphere \(S^2\).
A normalized state can be parametrized as
\begin{equation}
  |\psi(\theta,\phi)\rangle
  = \cos\frac{\theta}{2}\,|\uparrow\rangle
  + e^{i\phi}\sin\frac{\theta}{2}\,|\downarrow\rangle,
\end{equation}
with $\theta\in[0,\pi],\;\phi\in[0,2\pi)$.
The Fubini-Study line element reduces to the standard round metric,
\begin{equation}
  ds^2_{\mathrm{FS}} = d\theta^2 + \sin^2\theta\,d\phi^2,
\end{equation}
so that \(g_{\theta\theta}=1\), \(g_{\phi\phi}=\sin^2\theta\), and
\(\sqrt{\det g} = \sin\theta\).

To mimic the role of a macroscopic functional, we pick a simple scalar
function on the sphere, for instance
\begin{equation}
  f(\theta,\phi) = \cos\theta,
\end{equation}
which coincides with the expectation value of \(\sigma^z\) in the state
\(|\psi(\theta,\phi)\rangle\). Its derivatives are
\begin{equation}
  \partial_\theta f = -\sin\theta,
  \qquad
  \partial_\phi f = 0.
\end{equation}
The components of the Fubini-Study gradient are
\begin{equation}
\begin{split}
      (\nabla^{\mathrm{FS}} f)^\theta &= g^{\theta\theta}\partial_\theta f
  = -\sin\theta,\\
  (\nabla^{\mathrm{FS}} f)^\phi &= g^{\phi\phi}\partial_\phi f = 0,
\end{split}
\end{equation}
and the squared norm is
\begin{equation}
\begin{split}
      \|\nabla^{\mathrm{FS}} f\|_{g_{\mathrm{FS}}}^2
  &= g_{\theta\theta}(\nabla f)^\theta(\nabla f)^\theta
  + g_{\phi\phi}(\nabla f)^\phi(\nabla f)^\phi
  \\
  &= \sin^2\theta.
\end{split}
\end{equation}
The vector field $\bm\xi$ has components
\begin{equation}
  \xi^\theta = \frac{-\sin\theta}{\sin^2\theta}
  = -\frac{1}{\sin\theta},\qquad
  \xi^\phi = 0.
\end{equation}
Using the divergence formula~\eqref{eq:div_general} with
\(\sqrt{\det g}=\sin\theta\), we obtain
\begin{equation}
\begin{split}
      \mathrm{div}\,\bm \xi
  = \frac{1}{\sin\theta}\,\partial_\theta\big(\sin\theta \xi^\theta\big)
    &+ \frac{1}{\sin\theta}\,\partial_\phi\big(\sin\theta \xi^\phi\big)
  \\
  &= \frac{1}{\sin\theta}\,\partial_\theta(-1) = 0.
\end{split}
\end{equation}
In this toy example, the divergence of \(\bm \xi\) vanishes identically:
the level sets of \(f(\theta,\phi)=\cos\theta\) (parallels on the sphere)
are such that the particular combination of normal derivatives encoded in
\(\bm \xi\) has zero divergence. This warm-up illustrates concretely how the
objects entering the microcanonical construction are computed from the metric
and a chosen macroscopic functional.

\subsection{Two-qubit Schmidt family and entanglement}
\label{subsec:two_qubit_weingarten}

We now turn to a genuinely bipartite example where the macroscopic functional
is the bipartite entanglement entropy. Consider two qubits,
\(\mathcal H_A\simeq\mathbb C^2\), \(\mathcal H_B\simeq\mathbb C^2\),
with \(\mathcal H=\mathcal H_A\otimes\mathcal H_B\). Any pure state can be brought to the Schmidt form by local unitaries.
\begin{equation}
  |\psi(\theta,\phi)\rangle
  = \cos\theta\,|00\rangle
  + e^{i\phi}\sin\theta\,|11\rangle,
  \label{eq:twoqubit_schmidt}
\end{equation}
with $0\le\theta\le\tfrac{\pi}{4},\;\phi\in[0,2\pi)$ and
where we restrict \(\theta\) to the interval \([0,\pi/4]\) to avoid redundant labeling of Schmidt coefficients.

The reduced density matrix of subsystem \(A\) is
\begin{equation}
  \rho_A(\theta)
  = \mathrm{Tr}_B\,|\psi(\theta,\phi)\rangle\langle\psi(\theta,\phi)|
  = \begin{pmatrix}
      \cos^2\theta & 0 \\
      0 & \sin^2\theta
    \end{pmatrix},
\end{equation}
and the entanglement entropy is
\begin{equation}
  E(\theta)
  = -\cos^2\theta\log\cos^2\theta
    -\sin^2\theta\log\sin^2\theta.
\end{equation}
This depends only on \(\theta\), as expected from local-unitary invariance.

\paragraph{Fubini-Study metric on the Schmidt family.}

We now compute the restriction of the Fubini-Study metric to the
two-dimensional manifold parametrized by \((\theta,\phi)\).
The derivatives of \(|\psi(\theta,\phi)\rangle\) are
\begin{align}
  \partial_\theta|\psi\rangle
  &= -\sin\theta\,|00\rangle + e^{i\phi}\cos\theta\,|11\rangle,\\
  \partial_\phi|\psi\rangle
  &= i\,e^{i\phi}\sin\theta\,|11\rangle.
\end{align}
One readily finds
\begin{align}
  \langle\psi|\partial_\theta\psi\rangle &= 0, \\
  \langle\psi|\partial_\phi\psi\rangle &= i\sin^2\theta,\\
  \langle\partial_\theta\psi|\partial_\theta\psi\rangle &= 1,\\
  \langle\partial_\phi\psi|\partial_\phi\psi\rangle &= \sin^2\theta,\\
  \mathrm{Re}\,\langle\partial_\theta\psi|\partial_\phi\psi\rangle &= 0.
\end{align}
Therefore
\begin{equation}
\begin{split}
      \langle d\psi|d\psi\rangle
  &= \langle\partial_\theta\psi|\partial_\theta\psi\rangle\,d\theta^2
    + \langle\partial_\phi\psi|\partial_\phi\psi\rangle\,d\phi^2\\
  &= d\theta^2 + \sin^2\theta\,d\phi^2,
\end{split}
\end{equation}
and
\begin{equation}
  |\langle\psi|d\psi\rangle|^2
  = |\langle\psi|\partial_\phi\psi\rangle|^2\,d\phi^2
  = \sin^4\theta\,d\phi^2.
\end{equation}
The Fubini-Study line element~\eqref{eq:fs_general} hence becomes
\begin{equation}
  ds^2_{\mathrm{FS}}
  = 4\big(\langle d\psi|d\psi\rangle
          -|\langle\psi|d\psi\rangle|^2\big)
  = 4\big(d\theta^2
          + \sin^2\theta\cos^2\theta\,d\phi^2\big).
\end{equation}
Thus the metric components on this two-dimensional manifold are
\begin{equation}
  g_{\theta\theta} = 4,\qquad
  g_{\phi\phi} = 4\sin^2\theta\cos^2\theta,\qquad
  g_{\theta\phi}=0,
\end{equation}
with inverse metric
\begin{equation}
  g^{\theta\theta} = \frac14,\qquad
  g^{\phi\phi} = \frac{1}{4\sin^2\theta\cos^2\theta}.
\end{equation}
The determinant is
\(\det g = 16\sin^2\theta\cos^2\theta\), so that
\(\sqrt{\det g} = 4\sin\theta\cos\theta\) in the interval
\(\theta\in(0,\tfrac{\pi}{2})\).

\paragraph{Gradient and Weingarten trace.}

The derivative of the entanglement entropy with respect to \(\theta\) is
\begin{equation}
  \partial_\theta E
  = 4\sin\theta\cos\theta\,
    \log\big(\cot\theta\big),
\end{equation}
while \(\partial_\phi E=0\). The components of the Fubini
–Study gradient are therefore
\begin{equation}
\begin{split}
  (\nabla^{\mathrm{FS}}E)^\theta
  &= g^{\theta\theta}\partial_\theta E
  = \sin\theta\cos\theta\,\log(\cot\theta),\\
  (\nabla^{\mathrm{FS}}E)^\phi &= 0.
\end{split}
\end{equation}
The squared norm is
\begin{equation}
\begin{split}
      \|\nabla^{\mathrm{FS}}E\|_{g_{\mathrm{FS}}}^2
  &= g_{\theta\theta}(\nabla E)^\theta(\nabla E)^\theta
  \\
  &= 4\big(\sin\theta\cos\theta\,\log(\cot\theta)\big)^2.
\end{split}
\end{equation}
In this case, the vector field
$\bm \xi$ has components
\begin{equation}
\begin{split}
      \xi^\theta
  &= \frac{\sin\theta\cos\theta\log(\cot\theta)}
         {4\big(\sin\theta\cos\theta\log(\cot\theta)\big)^2}
  \\
  &= \frac{1}{4\sin\theta\cos\theta\,\log(\cot\theta)},\\
  \xi^\phi &= 0.
\end{split}
\end{equation}

Using the general divergence formula~\eqref{eq:div_general} with
\(\sqrt{\det g} = 4\sin\theta\cos\theta\), we find
\begin{equation}
\begin{split}
      \mathrm{div}\,\bm \xi
  &= \frac{1}{\sqrt{\det g}}\,
    \partial_\theta\big(\sqrt{\det g}\,\xi^\theta\big)\\
  &= \frac{1}{4\sin\theta\cos\theta}\,
    \partial_\theta\left(
      \frac{4\sin\theta\cos\theta}{4\sin\theta\cos\theta\,\log(\cot\theta)}
    \right)
  \\
  &= \frac{1}{4\sin\theta\cos\theta}\,
    \partial_\theta\left(\frac{1}{\log(\cot\theta)}\right).
\end{split}
\end{equation}
Since
\begin{equation}
  \partial_\theta\log(\cot\theta)
  = \frac{1}{\cot\theta}\,\partial_\theta(\cot\theta)
  = -\frac{1}{\sin\theta\cos\theta},
\end{equation}
we obtain
\begin{equation}
\begin{split}
      \partial_\theta\left(\frac{1}{\log(\cot\theta)}\right)
  &= -\frac{1}{\log^2(\cot\theta)}\,
    \partial_\theta\log(\cot\theta)
  \\
  &= \frac{1}{\sin\theta\cos\theta\,\log^2(\cot\theta)}.
\end{split}
\end{equation}
Substituting into the divergence gives
\begin{equation}
\begin{split}
  \mathrm{div}\,\bm \xi
  &= \frac{1}{4\sin\theta\cos\theta}\,
    \frac{1}{\sin\theta\cos\theta\,\log^2(\cot\theta)}
  \\
  &= \frac{1}{
      4\,\sin^2\theta\cos^2\theta\,\log^2(\cot\theta)
    }.
\end{split}
  \label{eq:twoqubit_divX}
\end{equation}

In this two-qubit Schmidt family the divergence~\eqref{eq:twoqubit_divX} provides
an explicit expression for the combination of normal derivatives that, in the
microcanonical construction, plays the role of the trace of the Weingarten
operator for the constant-entanglement level sets \(E(\theta)=\text{const}\).
Although the full constant-entanglement manifolds in \(\mathcal P(\mathcal H)
\simeq \mathbb{CP}^3\) are four-dimensional (real), this reduced calculation
already shows how the Fubini-Study geometry and the entanglement functional
combine to produce a nontrivial geometric invariant through the divergence
of \(\nabla^{\mathrm{FS}}E/\|\nabla^{\mathrm{FS}}E\|^2\).

This example can be viewed as the simplest nontrivial illustration of the general
prescription: given a macroscopic functional \(E\) on projective Hilbert space,
one computes the Fubini-Study gradient, constructs the vector field
\(\bm \xi = \nabla^{\mathrm{FS}}E/\|\nabla^{\mathrm{FS}}E\|^2\), and evaluates
its divergence to obtain the trace of the Weingarten operator in the chosen
convention. In higher-dimensional many-body systems, the structure is analogous,
but the explicit expressions become more involved.

\subsection{Microcanonical measure on the level set and geometric entropy}

For fixed \(\theta\), the set of states \(|\psi(\theta,\phi)\rangle\) with
\(\phi\in[0,2\pi)\) forms a one-dimensional submanifold of constant entanglement
\(E(\theta)\); in this reduced manifold it plays the role of the
constant-\(E\) hypersurface \(\Sigma_e\). Its Fubini-Study line element at
fixed \(\theta\) is
\begin{equation}
  ds^2_{\mathrm{FS}}\big|_{\theta=\text{const}}
  = 4\sin^2\theta\cos^2\theta\,d\phi^2,
\end{equation}
so that the induced Fubini-Study measure on the level set is
\begin{equation}
  d\sigma_{\mathrm{FS}}
  = \sqrt{g_{\phi\phi}}\,d\phi
  = 2\,\sin\theta\cos\theta\,d\phi.
\end{equation}
The microcanonical measure entering the density of states is
\begin{equation}
\begin{split}
      d\mu_e
  = \frac{d\sigma_{\mathrm{FS}}}{\|\nabla^{\mathrm{FS}}E\|}
  &= \frac{2\sin\theta\cos\theta\,d\phi}{
      2\sin\theta\cos\theta\,|\log(\cot\theta)|
    }
  \\
  &= \frac{d\phi}{|\log(\cot\theta)|}.
\end{split}
\end{equation}
The density of states at entanglement \(e=E(\theta)\) in this reduced
description is therefore
\begin{equation}
  \omega(e)
  = \int_{\Sigma_e} d\mu_e
  = \int_0^{2\pi}\frac{d\phi}{|\log(\cot\theta)|}
  = \frac{2\pi}{|\log(\cot\theta)|},
\end{equation}
and the corresponding geometric entanglement entropy is
\begin{equation}
  S_{\mathrm{geo}}(e)
  = \log\omega(e)
  = \log(2\pi) - \log|\log(\cot\theta)|,
  \label{eq:twoqubit_Sgeo_theta}
\end{equation}
with $e = E(\theta)$
\begin{figure*}
    \centering
    \includegraphics[width=0.8\linewidth]{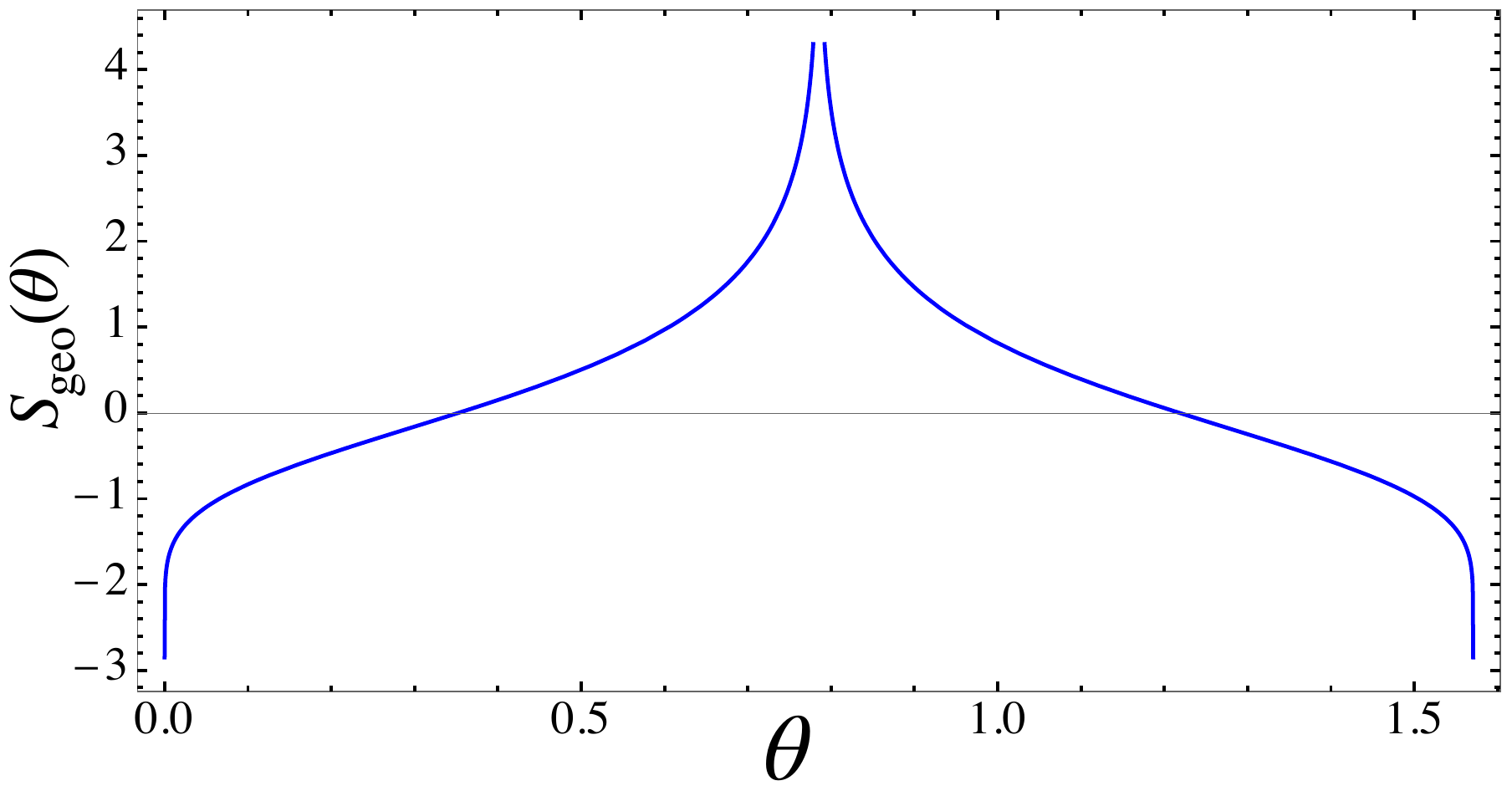}
    \caption{Geometric entropy $S_{\mathrm{geo}}(\theta)$ for a pair of qubits,
    as a function of the Schmidt angle $\theta$ in the parametrization
    $|\psi(\theta)\rangle=\cos\theta\,|00\rangle+\sin\theta\,|11\rangle$.
    The entropy diverges to $-\infty$ at $\theta\to 0,\pi/2$, corresponding
    to almost product states, showing that such low-entanglement states occupy
    a negligible Fubini-Study volume. The cusp at $\theta=\pi/4$ signals a
    strong concentration of volume around nearly maximally entangled states;
    the symmetry about $\pi/4$ reflects the degeneracy
    $\theta\leftrightarrow \tfrac{\pi}{2}-\theta$ of the Schmidt spectrum.}
    \label{fig:entropy-geo}
\end{figure*}

Fig.~\ref{fig:entropy-geo} shows that, in the two--qubit case, the geometric $S_{\mathrm{geo}}$ diverges to $-\infty$
at the endpoints $\theta\to 0,\pi/2$, where the entanglement vanishes and the states are (almost) product: the Fubini-Study volume associated with very small
entanglement is therefore negligible. In contrast, at $\theta=\pi/4$, corresponding
to a maximally entangled Bell state, $S_{\mathrm{geo}}$ develops a sharp cusp, indicating a strong concentration of volume around states with almost maximal
entanglement. The left--right symmetry reflects the fact that $\theta$ and $\tfrac{\pi}{2}-\theta$ generate the same Schmidt spectrum and hence the same entanglement. This simple example already captures the main message of the
construction: in the natural geometry of state space, almost separable pure states occupy an exceedingly small region, whereas nearly maximally entangled states are
geometrically typical.

\subsection{Average of \texorpdfstring{$\mathrm{Tr}\,W$}{Tr W} and \texorpdfstring{$\partial_e S_{\mathrm{geo}}$}{dS/de}}

The microcanonical average of the Weingarten trace at fixed entanglement is
defined as
\begin{equation}
  \big\langle \mathrm{Tr}\,W_{\bm\xi} \big\rangle_e
  = \frac{1}{\omega(e)}\int_{\Sigma_e}
    \mathrm{Tr}\,W_{\bm\xi}\,\frac{d\sigma_{\mathrm{FS}}}{\|\nabla^{\mathrm{FS}}E\|}.
\end{equation}
In the two-qubit Schmidt family, \(\mathrm{Tr}\,W_{\bm\xi}\) depends only on \(\theta\) and is constant along the orbit parameterized by
\(\phi\). Therefore
\begin{equation}
\begin{split}
      \int_{\Sigma_e}
    \mathrm{Tr}\,W_{\bm\xi}\,\frac{d\sigma_{\mathrm{FS}}}{\|\nabla^{\mathrm{FS}}E\|}
  &=\\
  &=\mathrm{Tr}\,W_{\bm\xi}(\theta)\int_0^{2\pi}\frac{d\phi}{|\log(\cot\theta)|}
  \\
  &= \mathrm{Tr}\,W_{\bm\xi}(\theta)\,\omega(e),
\end{split}
\end{equation}
and the average reduces to
\begin{equation}
  \big\langle \mathrm{Tr}\,W_{\bm\xi} \big\rangle_e
  = \mathrm{Tr}\,W_{\bm\xi}(\theta)
  = \frac{1}{
      4\,\sin^2\theta\cos^2\theta\,\log^2(\cot\theta)
    }.
\end{equation}

We now show explicitly that this coincides with \(\partial_e S_{\mathrm{geo}}(e)\).

From~\eqref{eq:twoqubit_Sgeo_theta} we have
\begin{equation}
\begin{split}
      \frac{d S_{\mathrm{geo}}}{d\theta}
  &= -\frac{1}{\log(\cot\theta)}\,
    \frac{d}{d\theta}\log(\cot\theta)
  \\
  &= -\frac{1}{\log(\cot\theta)}\,
    \left(-\frac{1}{\sin\theta\cos\theta}\right)
  \\
  &= \frac{1}{\sin\theta\cos\theta\,\log^2(\cot\theta)}.
\end{split}
\end{equation}
On the other hand, differentiating \(E(\theta)\) gives
\begin{equation}
  \frac{dE}{d\theta}
  = 4\sin\theta\cos\theta\,\log(\cot\theta).
\end{equation}
Hence, by the chain rule,
\begin{equation}
\begin{split}
  \frac{\partial S_{\mathrm{geo}}}{\partial e}
  &= \frac{d S_{\mathrm{geo}}/d\theta}{dE/d\theta}
  \\
  &= 
      \dfrac{1}{\sin\theta\cos\theta\,\log^2(\cot\theta)}\frac{1
    }{
      4\sin\theta\cos\theta\,\log(\cot\theta)
    }\\
  &= \frac{1}{
      4\,\sin^2\theta\cos^2\theta\,\log^2(\cot\theta)
    }.
\end{split}
\end{equation}
Comparing with~\eqref{eq:twoqubit_divX} we obtain
\begin{equation}
  \frac{\partial S_{\mathrm{geo}}}{\partial e}
  = \big\langle \mathrm{Tr}\,W \big\rangle_e,
\end{equation}
which is precisely the relation
\begin{equation}
  \partial_e S_{\mathrm{geo}}(e)
  = \frac{1}{\omega(e)}\int_{\Sigma_e}
    \mathrm{Tr}\,W\,\frac{d\sigma_{\mathrm{FS}}}{\|\nabla^{\mathrm{FS}}E\|}
\end{equation}
in this simple two-qubit example.

This explicit calculation shows, in a completely controllable setting, how the
geometric objects entering the microcanonical construction---the Fubini–Study metric---the entanglement functional \(E\), the Weingarten trace, and the
microcanonical measure \(d\sigma_{\mathrm{FS}}/\|\nabla^{\mathrm{FS}}E\|\)—
combine to produce the expected relation between the geometric entropy and the mean curvature of constant-entanglement manifolds.

\begin{figure*}
    \centering
    \includegraphics[width=0.8\textwidth]{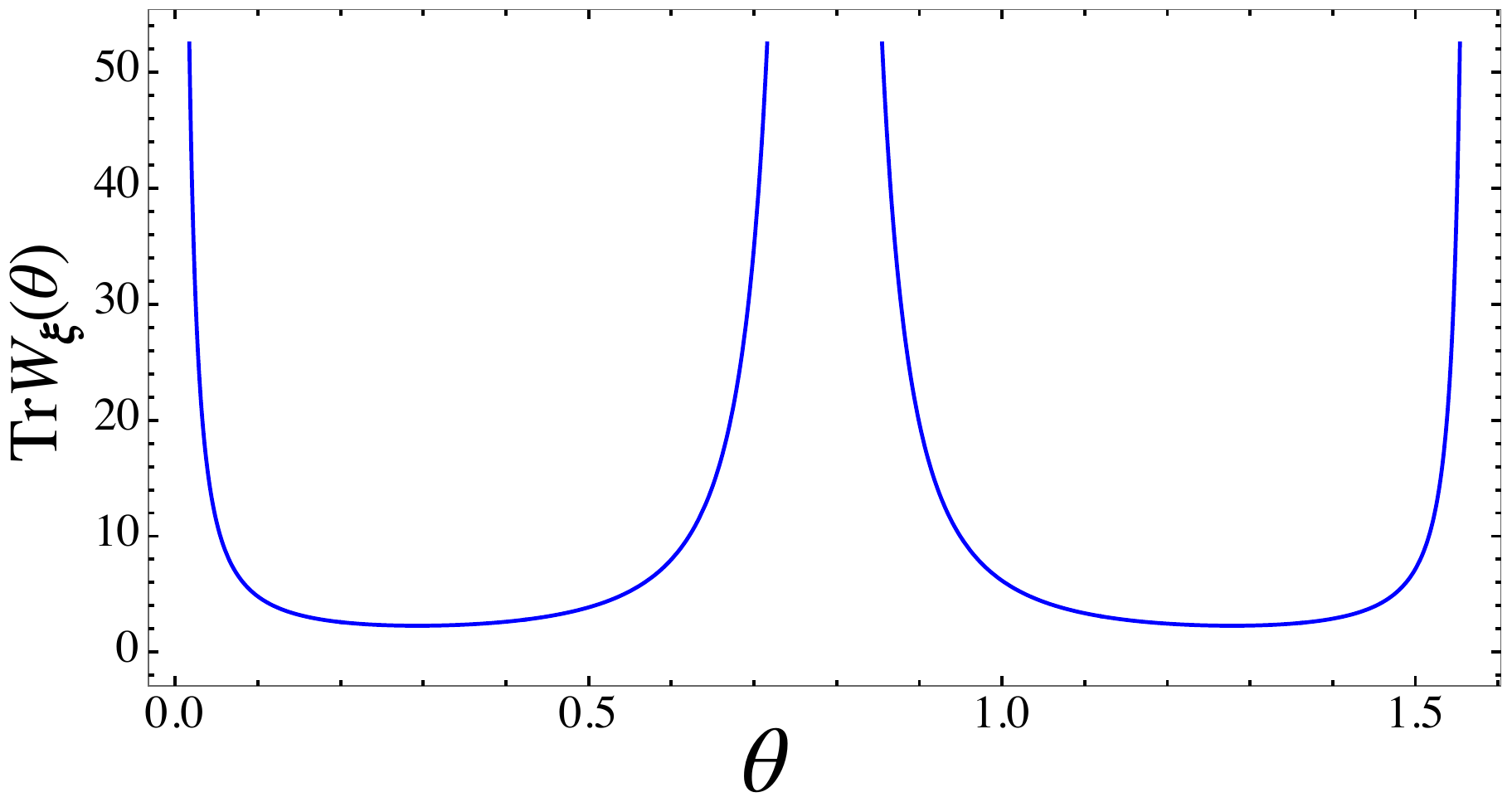}
    \caption{Trace of the Weingarten operator,
    $\mathrm{Tr}\,W(\theta)=\mathrm{div}\,\bm \xi$, for the two--qubit
    Schmidt family
    $|\psi(\theta)\rangle=\cos\theta\,|00\rangle+\sin\theta\,|11\rangle$.
    The trace is positive and diverges at $\theta\to 0,\pi/2$ (almost product
    states) and at $\theta=\pi/4$ (maximally entangled state), indicating
    very strong extrinsic curvature of the constant–entanglement hypersurfaces
    in the Fubini-Study geometry near both separable and maximally entangled
    configurations. For intermediate $\theta$ the curvature is finite and
    relatively small, corresponding to a slower variation of the geometric
    entropy with entanglement.}
    \label{fig:twoqubit_TrW}
\end{figure*}

Fig.~\ref{fig:twoqubit_TrW} shows
$\mathrm{Tr}\,W(\theta)$ as a function of the Schmidt angle. The trace is strictly positive and diverges at the endpoints $\theta\to 0,\pi/2$, where
the state becomes (almost) product and either $\sin\theta$ or $\cos\theta$ vanishes; it also diverges at $\theta=\pi/4$, where the entanglement is maximal and $\log(\cot\theta)\to 0$. These divergences signal that the
constant–entanglement hypersurfaces are extremely curved, in the Fubini-Study geometry, both near separable states and near the maximally entangled state.
In contrast, for intermediate values of $\theta$ the Weingarten trace remains finite and relatively small: in these regions the variation of the geometric entropy with entanglement is comparatively mild and the constant–$E$ manifolds are geometrically flatter. Since in this example
$\partial_e S_{\mathrm{geo}}(e) = \mathrm{Tr}\,W(\theta(e))$, the plot
directly represents the slope of the geometric entropy as a function of entanglement.

\section{Discussion and outlook}
\label{sec:discussion}

We have introduced a geometric framework in which bipartite entanglement is treated
as a macroscopic functional on projective Hilbert space, endowed with the
Fubini-Study metric. This viewpoint leads naturally to the notion of a
\emph{geometric entanglement entropy} $S_{\mathrm{geo}}(e)$, defined as the
logarithm of the Fubini-Study volume of the manifold of pure states with fixed
entanglement $e$. Conceptually, $S_{\mathrm{geo}}(e)$ measures how abundant or rare
different entanglement regimes are in the space of all pure states, and plays the
role of a microcanonical entropy in entanglement space.

Several directions for future work are natural. On the conceptual side, one would like to characterize more explicitly the curvature invariants entering
$\Omega_{\mathrm{FS}}(e)$ and relate them to known structures in quantum information, such as local unitary orbits~\cite{bengtsson_zyczkowski},
entanglement polytopes~\cite{entanglement_polytopes_review}, and majorization relations~\cite{nielsen_majorization,horodecki_review}. On the computational side, it would be interesting to obtain explicit or approximate forms of
$S_{\mathrm{geo}}(e)$ for low-dimensional systems (two or three qubits, qutrit systems) and to compare them with numerical estimates obtained by sampling random pure states.

In many-body systems, one could study how $S_{\mathrm{geo}}(e)$ scales with
system size for different bipartitions and whether distinct curvature regimes
of $\Sigma_e$ can be associated with familiar notions such as area-law versus
volume-law entanglement~\cite{Eiser2008entanglement,EisertReview} or with
transitions in entanglement structures observed in random circuits and
measurement-induced transitions~\cite{nahum2017quantum,li2018quantum,skinner2019measurement}.
Finally, one might consider extensions of the present framework to mixed states, possibly using generalized Riemannian metrics on the state space of density
matrices~\cite{petz_monotone,petz_sudar} and entanglement measures beyond the von Neumann entropy.

The broader message is that the projective Hilbert space, endowed with the Fubini-Study metric, contains a rich geometrical encoding of the entanglement structure, which can be accessed without committing to any particular dynamics. The geometric entanglement entropy $S_{\mathrm{geo}}(e)$ provides a compact way of organizing this information and may offer a useful bridge between the geometry of quantum state space and the resource-theoretic perspective on entanglement.


\begin{thebibliography}{99}

\bibitem{nielsen_chuang}
M.~A.~Nielsen and I.~L.~Chuang,
\textit{Quantum Computation and Quantum Information}
(Cambridge University Press, Cambridge, 2000).

\bibitem{horodecki_review}
R.~Horodecki, P.~Horodecki, M.~Horodecki, and K.~Horodecki,
\textit{Quantum entanglement},
Reviews of Modern Physics \textbf{81}, 865--942 (2009).

\bibitem{page1993typical}
D.~N.~Page,
\textit{Average entropy of a subsystem},
Physical Review Letters \textbf{71}, 1291--1294 (1993).

\bibitem{foong_kanno}
S.~K.~Foong and S.~Kanno,
\textit{Proof of Page's conjecture on the average entropy of a subsystem},
Physical Review Letters \textbf{72}, 1148--1151 (1994).

\bibitem{typical_entanglement_review}
F.~Deelan Cunden, P.~Facchi, G.~Florio, and S.~Pascazio,
\textit{Typical entanglement},
The European Physical Journal Plus \textbf{128}, 48 (2013).


\bibitem{popescu2006entanglement}
S.~Popescu, A.~J.~Short, and A.~Winter,
\textit{Entanglement and the foundations of statistical mechanics},
Nature Physics \textbf{2}, 754--758 (2006).

\bibitem{hayden2006aspects}
P.~Hayden, D.~W.~Leung, and A.~Winter,
\textit{Aspects of Generic Entanglement},
Communications in Mathematical Physics \textbf{265}, 95--117 (2006).

\bibitem{Eiser2008entanglement}
L.~Eiser, R.~Fazio, A.~Osterloh, and V.~Vedral,
\textit{Entanglement in many-body systems},
Reviews of Modern Physics \textbf{80}, 517--576 (2008).

\bibitem{eisert2010colloquium}
J.~Eisert, M.~Cramer, and M.~B.~Plenio,
\textit{Colloquium: Area laws for the entanglement entropy},
Reviews of Modern Physics \textbf{82}, 277--306 (2010).

\bibitem{calabrese2009entanglement}
P.~Calabrese, J.~Cardy, and B.~Doyon,
\textit{Entanglement entropy in extended quantum systems},
Journal of Physics A: Mathematical and Theoretical \textbf{42}, 500301 (2009).

\bibitem{bengtsson_zyczkowski}
I.~Bengtsson and K.~{\.Z}yczkowski,
\textit{Geometry of Quantum States: An Introduction to Quantum Entanglement},
2nd ed.~(Cambridge University Press, Cambridge, 2017).

\bibitem{provost1980riemannian}
J.~P.~Provost and G.~Vallee,
\textit{Riemannian structure on manifolds of quantum states},
Communications in Mathematical Physics \textbf{76}, 289--301 (1980).


\bibitem{brody2001geometric}
D.~C.~Brody and L.~P.~Hughston,
\textit{Geometric quantum mechanics},
Journal of Geometry and Physics \textbf{38}, 19--53 (2001).

\bibitem{di2022geometrictheory}
L.~Di~Cairano,
\textit{The Geometric Theory of Phase Transitions},
Journal of Physics A: Mathematical and Theoretical \textbf{55}, 27LT01 (2022).

\bibitem{federer_gmt}
H.~Federer,
\textit{Geometric Measure Theory}
(Springer, Berlin, 1969).

\bibitem{vedralReview}
V.~Vedral,
\textit{The role of relative entropy in quantum information theory},
Reviews of Modern Physics \textbf{74}, 197--234 (2002).

\bibitem{nielsen_majorization}
M.~A.~Nielsen,
\textit{Conditions for a class of entanglement transformations},
Physical Review Letters \textbf{83}, 436--439 (1999).


\bibitem{li2018quantum}
Y.~Li, X.~Chen, and M.~P.~A.~Fisher,
\textit{Quantum Zeno effect and the many-body entanglement transition},
Physical Review B \textbf{98}, 205136 (2018).

\bibitem{petz_monotone}
D.~Petz,
\textit{Monotone metrics on matrix spaces},
Linear Algebra and its Applications \textbf{244}, 81--96 (1996).

\bibitem{petz_sudar}
D.~Petz and C.~Sud{\'a}r,
\textit{Geometries of quantum states},
Journal of Mathematical Physics \textbf{37}, 2662--2673 (1996).

\bibitem{franzosi2018microcanonical}
R.~Franzosi,
\textit{Microcanonical entropy for classical systems},
Physica A: Statistical Mechanics and its Applications \textbf{494}, 302--307 (2018).

\bibitem{franzosi2019microcanonical}
R.~Franzosi,
\textit{A microcanonical entropy correcting finite-size effects in small systems},
Journal of Statistical Mechanics: Theory and Experiment \textbf{2019}, 083204 (2019).

\bibitem{di2021topology}
L.~Di~Cairano, M.~Gori, and M.~Pettini,
\textit{Topology and Phase Transitions: A First Analytical Step towards the Definition of Sufficient Conditions},
Entropy \textbf{23}, 1414 (2021).

\bibitem{hirsch2012differential}
M.~W.~Hirsch,
\textit{Differential Topology}
(Springer Science \& Business Media, 2012).

\bibitem{franzosi2000topology}
R.~Franzosi, M.~Pettini, and L.~Spinelli,
\textit{Topology and phase transitions: Paradigmatic evidence},
Physical Review Letters \textbf{84}, 2774 (2000).

\bibitem{wei_geometric_entanglement}
T.-C.~Wei and P.~M.~Goldbart,
\textit{Geometric measure of entanglement and applications to bipartite and multipartite quantum states},
Physical Review A \textbf{68}, 042307 (2003).

\bibitem{orus_geometric_entanglement}
R.~Or\'us and T.-C.~Wei,
\textit{Geometric entanglement in topologically ordered states},
Physical Review B \textbf{88}, 075117 (2013).

\bibitem{orus_review}
R.~Or\'us,
\textit{A practical introduction to tensor networks: Matrix product states and projected entangled pair states},
Annals of Physics \textbf{349}, 117--158 (2014).


\bibitem{leinaas_geometrical_aspects}
J.~M.~Leinaas, J.~Myrheim, and E.~Ovrum,
\textit{Geometrical aspects of entanglement},
Physical Review A \textbf{74}, 012313 (2006).

\bibitem{frydryszak_geometric_measure}
A.~M.~Frydryszak and V.~M.~Tkachuk,
\textit{Quantifying geometric measure of entanglement by mean value of spin and spin correlations for pure and mixed states},
European Physical Journal D \textbf{71}, 233 (2017).

\bibitem{vesperini_geometric_entanglement}
A.~Vesperini, G.~Bel{-}Hadj{-}Aissa, L.~Capra, and R.~Franzosi,
\textit{Unveiling the geometric meaning of quantum entanglement: Discrete and continuous variable systems},
Frontiers of Physics \textbf{19}, 51204 (2024).

\bibitem{anza_maximum_geometric_entropy}
F.~Anza and J.~P.~Crutchfield,
\textit{Maximum Geometric Quantum Entropy},
Entropy \textbf{26}, 225 (2024).

\bibitem{qid_geometric_entropy}
O.~A.~Castro{-}Alvaredo and B.~Doyon,
\textit{Entanglement in permutation symmetric states, fractal dimensions, and geometric quantum mechanics},
Journal of Statistical Mechanics: Theory and Experiment \textbf{2013}, P02016 (2013).

\bibitem{entanglement_polytopes_review}
M.~Walter, B.~Doran, D.~Gross, and M.~Christandl,
\textit{Entanglement Polytopes: Multiparticle Entanglement from Single-Particle Information},
Science \textbf{340}, 1205--1208 (2013).

\bibitem{bel2020geometrical}
G.~Bel-Hadj-Aissa, M.~Gori, V.~Penna, G.~Pettini, and R.~Franzosi,
\textit{Geometrical aspects in the analysis of microcanonical phase-transitions},
Entropy \textbf{22}, 380 (2020).

\bibitem{gori_configurational}
M.~Gori,
\textit{Configurational microcanonical statistical mechanics from Riemannian geometry of equipotential level sets},
arXiv:2205.14536 (2022).

\bibitem{gori2022topological}
M.~Gori, R.~Franzosi, G.~Pettini, and M.~Pettini,
\textit{Topological theory of phase transitions},
Journal of Physics A: Mathematical and Theoretical \textbf{55}, 375002 (2022).

\bibitem{zhou2013simple}
Y.~Zhou,
\textit{A simple formula for scalar curvature of level sets in Euclidean spaces},
arXiv:1301.2202 (2013).


\bibitem{nahum2017quantum}
A.~Nahum, J.~Ruhman, S.~Vijay, and J.~Haah,
\textit{Quantum Entanglement Growth Under Random Unitary Dynamics},
Physical Review X \textbf{7}, 031016 (2017).

\bibitem{skinner2019measurement}
B.~Skinner, J.~Ruhman, and A.~Nahum,
\textit{Measurement-Induced Phase Transitions in the Dynamics of Entanglement},
Physical Review X \textbf{9}, 031009 (2019).

\bibitem{cocchiarella2020entanglement}
D.~Cocchiarella, S.~Scali, S.~Ribisi, B.~Nardi, G.~Bel-Hadj-Aissa, and R.~Franzosi,
\textit{Entanglement distance for arbitrary m-qudit hybrid systems},
Physical Review A \textbf{101}, 042129 (2020).

\bibitem{vesperini2023entanglement}
A.~Vesperini, G.~Bel-Hadj-Aissa, and R.~Franzosi,
\textit{Entanglement and quantum correlation measures for quantum multipartite mixed states},
Scientific Reports \textbf{13}, 2852 (2023).

\bibitem{vesperini2024entanglement}
A.~Vesperini and R.~Franzosi,
\textit{Entanglement, quantum correlators, and connectivity in graph states},
Advanced Quantum Technologies \textbf{7}, 2300264 (2024).

\bibitem{de2025entanglement}
L.~De~Simone and R.~Franzosi,
\textit{Entanglement in quantum systems based on directed graphs},
Journal of Physics A: Mathematical and Theoretical \textbf{58}, 415302 (2025).

\end{thebibliography}
\end{document}